\documentclass[aps,jmp, amsmath,amssymb,reprint]{revtex4-1}

\usepackage{graphicx}
\usepackage{dcolumn}
\usepackage{bm}
\usepackage{color}

\begin{document}


\title[]{Computing the dielectric constant of liquid water at constant dielectric displacement}

\author{Chao Zhang}
\email{cz302@cam.ac.uk}
\author{Michiel Sprik}%
\affiliation{Department of Chemistry, University of Cambridge,
  Lensfield Rd, Cambridge CB2 1EW, United Kingdom
}%


\date{\today}

\begin{abstract}
The static dielectric constant of liquid water is computed using classical force field based molecular dynamics simulation at fixed electric displacement $\mathbf{D}$. The method to constrain the electric displacement is the finite temperature classical variant of the constant-$\mathbf{D}$ method developed by Stengel, Spaldin and Vanderbilt [Nat. Phys. {\bf 5}, 304, (2009)]. There is also a modification of this scheme imposing fixed values of the macroscopic field $\mathbf{E}$. The method is applied to the popular SPC/E model of liquid water. We compare four different estimates of the dielectric constant, two obtained from fluctuations of the polarization at $\mathbf{D}=0$ and $\mathbf{E}=0$ and two from the variation of polarization with finite $\mathbf{D}$ and $\mathbf{E}$. It is found that all four estimates agree when properly converged. The computational effort to achieve convergence varies however, with constant $\mathbf{D}$ calculations being substantially more efficient. We attribute this difference to the much shorter relaxation time of longitudinal polarization compared to transverse polarization accelerating constant $\mathbf{D}$ calculations.      

%
\end{abstract}

\maketitle

%

\section{Introduction}

The static dielectric constant of model polar liquids is usually computed from polarization fluctuations applying the linear response relations of Kirkwood-Fr\"ohlich theory~\cite{Kirkwood:1939br, frohlich1958theory}. These calculations are carried out in periodic molecular dynamics (MD) cells treating long range electrostatic interactions using Ewald summation\cite{DeLeeuw:1980rs80a}.  This is an expensive calculation. Indeed, it is commonly acknowledged that for highly polarizable liquids such as water simulations at time-scales of nanoseconds are necessary to converge the fluctuation estimate for the dielectric constant.\cite{DeLeeuw:1980rs80a,DeLeeuw:1980rs80b,DeLeeuw:1986gs,Neumann:1983gy, Neumann:1983gz, Neumann:1983kg, Neumann:1984mp, Neumann:1985bq, Neumann:1986km, Patey:1982mp, Caillol:1989jcp, REDDY:1989jn, WATANABE:1989bu, Sprik:1991fx, Rick:1994km, vanderSpoel:1998gb, Hunenberger:1998el, Lamoureux:2006ih, Aragones:2011dp, Templeton:2013jcp, Zhang:2014jq, Braun:2014fs, Elton:2014jcp, Kolafa:2014jctc}. These time scales are accessible for classical force field based MD simulation. This is, however, a major challenge for MD simulation with forces calculated ``on-the-fly'' using electronic structure calculation methods, such as density functional theory~\cite{Umari:2004cpl,Sharma:2007prl,Pan:2013kb}. 

The high costs of the Kirkwood-Fr\"ohlich scheme is a strong motivation for the development for more efficient alternatives. Boundary conditions can have a drastic effect on polarization fluctuations which has led to the search for optimal boundary conditions\cite{Neumann:1983gz,Neumann:1984mp,Hunenberger:1998el,Kolafa:2014jctc}. The use of finite field methods is another option that has been investigated\cite{Yeh:1999hb,Kolafa:2014jctc,vanGunsteren:2011jctc}. The rationale here is, of course, that converging an estimate of polarization should be quicker than converging its fluctuations. However it was found, in particular in the case of water, that the response of the polarization  is non-linear for already moderate field strength requiring a careful extrapolation to zero field. As a result finite field calculations of the dielectric constant are in practice not that much cheaper than the computation from polarization fluctuations at zero field. 

A change of boundary conditions not only affects the size of polarization fluctuations but also the time scale. It has been shown that the standard Ewald summation method corresponds to constraining the average macroscopic field to zero\cite{DeLeeuw:1980rs80a,Neumann:1983gy}. The static dielectric constant under these conditions is dominated by fluctuations of the transverse polarization\cite{Fulton:1975ii,Madden:1984cp,Kivelson:1989fj}. These are the slow modes. The relaxation time $\tau_{L}$  of longitudinal modes (in the $\mathbf{k}=0$ limit) is significantly faster compared to the relaxation time $\tau_T$ of transverse modes. The ratio according to Debye theory\cite{frohlich1958theory} is  $\tau_{T}/\tau_{L} = \epsilon_0/\epsilon_{\infty}$. For  non-polarizable SPC models of liquid water ($\epsilon_{\infty} = 1$) this amounts almost to two orders of magnitude. 

The much faster relaxation of longitudinal fluctuations raises the question whether this can be exploited to accelerate the calculation of the static dielectric constant. In this paper we show that this can be achieved by changing the boundary conditions for Ewald summation from zero macroscopic field ($\mathbf{E}=0$) to zero dielectric displacement ($\mathbf{D}=0$). The method to compute the  total energy of periodic supercells under fixed $\mathbf{D}$ has been developed by Vanderbilt and coworkers. The key reference to this approach is the 2009 Nature Physics paper by Stengel, Spaldin and Vanderbilt\cite{Stengel:2009cd} which will be referred to as SSV (see also Ref.~\citenum{Stengel:2009prb}). The method is a recent spin-off of the modern theory of polarization developed by Vanderbilt and Resta during the 90's\cite{King-Smith:1993prb,Resta:1994rmp,Resta:2007ch}. The modern theory of polarization caused a revolution in theoretical and computational solid state physics making it possible, for the first time, to investigate the electric equation of state of ferroelectric systems. The initial approach was to compute the total energy for fixed values of the polarization and to determine the electric field from the derivative\cite{Sai:2004prb,Dieguez:2006prl}. This was subsequently changed to a scheme using directly the macroscopic electric field $\mathbf{E}$ or the electric displacement field $\mathbf{D}$  as the control variable, which has both computational and conceptual advantages\cite{Stengel:2009cd,Stengel:2009prb}.   

The SSV constant $\mathbf{D}$ and the related finite $\mathbf{E}$ method are easy to implement in a classical force field code. The method retains the regular ``tin-foil'' Ewald sum for the calculation of electrostatic energy extending it with an electric term which depends on the polarization $\mathbf{P}$ and contains $\mathbf{E}$ or $\mathbf{D}$ as a parameter. The result is an extended hamiltonian replacing the original hamiltonian in the MD simulation. The present paper is a feasibility and validation study of this approach for a SPC/E model of liquid water~\cite{Berendsen:1987uu}. We verify that the dielectric constant obtained from polarization fluctuations under $\mathbf{D}=0$ conditions agrees with the value estimated from standard $\mathbf{E}=0$ calculations. These results are then also compared to dielectric constant estimates computed from the change of the expectation value of  $\mathbf{P}$ with applied $\mathbf{E}$ and $\mathbf{D}$ using the SSV hamiltonian in finite field mode.

The approach in the present paper has multiple parallels to similar work that has appeared in the literature. Most of this work also involves macroscopic polarization dependent energy terms extending the microscopic Coulomb interaction energy evaluated for an infinite periodic lattice of supercells using the Ewald summation method\cite{Ewald21}. These extensions are known as surface terms\cite{DeLeeuw:1980rs80a,DeLeeuw:1980rs80b,Redlack:1975jpcs,Smith:1981rsa,Smith:2008jcp,Caillol:1994jcp,Foulkes:1996prb,Kantorovich:1999jpcm,Ballenegger:2014jpc} or reaction fields\cite{Neumann:1983gy,Neumann:1983gz,Neumann:1983kg} and are of a form similar to the $\mathbf{D}=0$ limit of the SSV constant $\mathbf{D}$ polarization coupling term. What is unique about SSV theory is that the extended Ewald Hamiltonians are derived strictly complying with the rules of dielectric thermodynamics as set out by Landau and Lifshitz\cite{Landau:1960v8}. The focus on thermodynamics has certain advantages as already pointed by Aragones et al.\cite{Vega:2011prl}. The paper starts therefore with a fairly detailed outline of the finite temperature classical variant of the SSV approach (sections \ref{sec:ED} and \ref{sec:diel}) supplemented with three appendices with more formal theoretical considerations. Results are presented and discussed in section \ref{sec:results}. We conclude with a summary and outlook for future applications to interfaces.

\section{Finite E and D in extended systems} \label{sec:ED}

 \subsection{Constant E and D hamiltonians and thermodynamics} \label{sec:EDham}
 The theory behind the finite field method developed by SSV  is summarized in the supporting information of Ref.~\citenum{Stengel:2009cd}. The central quantity  is the  electric enthalpy functional.  The electric enthalpy of a system of volume $\mathit{\Omega}$ is written as
\begin{equation}
\label{fvdb}
\mathcal{F}\left( \mathbf{E}, v \right) = E_{\mathrm{KS}}(v)- \mathit{\Omega}  \,\mathbf{E} \cdot \mathbf{P}(v)
\end{equation}
 $E_{\mathrm{KS}}(v)$ is the Kohn-Sham total energy with $v$ denoting all microscopic degrees of freedom involved, ie the orbital coefficients specifying the one-electron orbitals and the positions of the ions.  $E_{\mathrm{KS}}(v)$ is obtained for given $v$  using the regular reciprocal space  methods of computational solid state physics explicitly excluding all $\mathbf{k}=0$ contributions.  $\mathbf{P}(v)$ is the macroscopic polarization density for the microscopic state specified by $v$.

 To compute the expectation value $\mathbf{P}$ of the polarization density the electronic structure $v$ of the system is determined by minimizing the electric enthalpy functional for fixed  $\mathbf{E}$
\begin{equation}
\label{fks}
F\left( \mathbf{E}\right) = \min_{v} \mathcal{F}\left( \mathbf{E}, v) \right) =
 \min_{v} \lbrack E_{\mathrm{KS}}(v) - \mathit{\Omega}  \,\mathbf{E} \cdot \mathbf{P}(v) \rbrack
\end{equation}
Taking the derivative gives the polarization
\begin{equation}
\label{dfde}
\frac{d F}{d \mathbf{E}} = -\mathit{\Omega} \mathbf{P} 
\end{equation}
Eq.~\ref{dfde} can be regarded as the electric equation of state for a uniform insulator.

Vanderbilts electric enthalpy  scheme can be readily adapted to classical force field based MD. The KS total energy   $E_{\mathrm{KS}}(v)$ in Eq.~\ref{fvdb} is replaced  by the Hamiltonian $H(v)$ of the system where $v$ is now the set of momenta and positions of the particles. 
\begin{equation}
\label{fhmd}
\mathcal{F}\left( \mathbf{E}, v \right) = H_{\mathrm{PBC}}(v)- \mathit{\Omega}  \,\mathbf{E} \cdot \mathbf{P}(v)
\end{equation}
We have appended a subscript PBC to the Hamiltonian as a reminder that the electrostatic energies and forces are computed using standard Ewald summation as applied also for the computation of the KS energy in Eq.~\ref{fvdb}. $H_{\mathrm{PBC}}(v)$ can be formally written as the sum of a term $H_{sr}$ describing the short range interactions and the reciprocal space representation of the electrostatic energy 
\begin{equation}
\label{hpbc}
H_{\mathrm{PBC}}(v) = H_{sr}(v) + 2 \pi \Omega \sum_{\mathbf{k} \neq 0} 
  \frac{\rho(\mathbf{k})^2}{k^2}
\end{equation}
where $\rho(\mathbf{k})$ is the Fourier transform of the atomic charge distribution (the SPC charges). Ewald summation is a computationally more efficient method to calculate this energy carrying out the summation partially in real space. The result corresponds to zero average electric field and potential, the so called tinfoil boundary conditions (no surface terms)\cite{DeLeeuw:1980rs80a}. The finite electric field is introduced as a parameter in the second term of Eq.~\ref{fhmd}.  

The equivalent of the electric enthalpy of Eq.~\ref{fks} is the free energy of the ensemble generated by the extended Hamiltonian $F\left( \mathbf{E}, v \right)$ of Eq.~\ref{fhmd}
\begin{equation}
\label{fmd}
F\left( \mathbf{E}\right) = - k_{\mathrm{B}}T \ln Z_E
\end{equation}
$Z_E $ is the electric field dependent partition function 
\begin{equation}
\label{zfmd}
Z_E = \int d\mathbf{p}^N d\mathbf{r}^N \exp \lbrack - \beta \left(
  H_{\mathrm{PBC}} - \mathit{\Omega}  \,\mathbf{E} \cdot \mathbf{P}(\mathbf{r}^N)\right)  \rbrack
\end{equation}
where $ H_{\mathrm{PBC}}$ is the Hamiltonian  of the periodic MD system in Eq.~\ref{fhmd}. The coordinate  and momentum arguments $v= \mathrm{r}^N ,\mathrm{p}^N$ were suppressed. $\beta = 1/k_{\mathrm{B}}T $ is the inverse temperature. The combinatorial prefactor $1/(h^{3N} N!)$ has been omitted. The derivative of $F\left( \mathbf{E}\right)$ of Eq.~\ref{fmd} again gives the polarization according to Eq.~\ref{dfde}.

Recently,  Stengel, Spaldin and Vanderbilt have modified the constant $\mathbf{E}$ to a constant $\mathbf{D}$ method~\cite{Stengel:2009cd}. They introduced a new functional, the electric internal energy functional $\mathcal{U}\left( \mathbf{D}, v \right)$.  Transposed to classical MD,  this functional is written as
 \begin{equation}
\label{uvdb}
\mathcal{U}\left( \mathbf{D}, v \right) = H_{\mathrm{PBC}}(v)+ \frac{\mathit{\Omega}}{8 \pi}
  \left( \mathbf{D} - 4 \pi \mathbf{P}(v) \right)^2
\end{equation}
The corresponding $\mathbf{D}$ dependent electric internal energy is again obtained from the partition function
\begin{equation}
\label{umd}
U\left( \mathbf{D}\right) = - k_{\mathrm{B}}T \ln Z_D
\end{equation}
with
\begin{equation}
\label{zumd}
Z_D = \int d\mathbf{p}^N d\mathbf{r}^N \exp \left[ - \beta \, \mathcal{U}(\mathrm{D}, v) \right]
\end{equation}
where as before  $v= \mathrm{r}^N ,\mathrm{p}^N$ . Note that $U(\mathbf{D})$  is still a (Helmholtz) free energy with respect to temperature, similar to $F(\mathbf{E})$.  Evaluating the derivative with respect the control variable, $\mathbf{D}$ in this case, we recover the macroscopic field 
\begin{equation}
\label{dudd}
 \frac{d U}{d \mathbf{D}} =\frac{\mathit{\Omega}}{4 \pi} \left(\mathbf{D}-4\pi \mathbf{P} \right)
= \frac{\mathit{\Omega}}{4 \pi} \mathbf{E}
\end{equation}
The second identity follows from 
\begin{equation}
\label{ddef}
\mathbf{D}=\mathbf{E}+4\pi\mathbf{P}
\end{equation}
which is the  fundamental relation of Maxwell theory defining the dielectric displacement.

That $U$ of Eq.~\ref{umd} is indeed the electric free energy becomes evident when Eq.~\ref{dudd} is substituted in the Maxwell field expression for electrical work (see Landau and Lifshitz\cite{Landau:1960v8})
\begin{equation}
\label{ework}
dW = \frac{\mathit{\Omega}}{4 \pi} \mathbf{E} \cdot d\mathbf{D} =dU
\end{equation}
The link to electrical work established in Eq.~\ref{ework} is crucial. It is the ultimate justification for identifying $\mathbf{E}$ in the electrical enthalpy of Eq.~\ref{fhmd} with the macroscopic field. As shown by SSV, the argument can be given a more formal thermodynamic basis in a Legendre transform framework. The derivation is repeated in Appendix \ref{sec:leg}.

\subsection{Parallel plate capacitor and hybrid boundary conditions} 
\label{sec:cap}
\begin{figure}
\includegraphics[width=0.95\columnwidth]{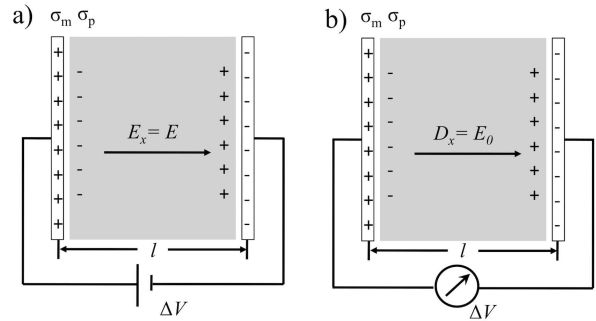}
\caption{\label{fig0} Parallel plate capacitor at a) constant electric field $E_x$ and b) at constant electric displacement $D_x$. $\sigma_m$ is the surface charge density on the metal electrode. $\sigma_p$ is the polarization surface charge density of the dielectric material ($\sigma_m > -\sigma_p > 0 $ in the picture). The electric field $E$ is determined by the net interface charge $\sigma_m + \sigma_p$ (see text).}
\end{figure}

The SSV Hamiltonians of Eqs.~\ref{fhmd} and \ref{uvdb} can be given more physical meaning when interpreted as a model of a macroscopic parallel plate capacitor\cite{Stengel:2009cd,Stengel:2009prb}. Fig.~\ref{fig0} shows a schematic picture of such a device as used in textbooks (see in particular Purcell\cite{Purcell:2011ca}).  We will assume that the normal to the electrodes is directed along the $x$ axis. The electrodes are a distance $l$ apart. The charge density on the left electrode is $\sigma_m$. Without dielectric material between the plates the electric field $E_0$ generated by this charge density is $E_0 = 4 \pi \sigma_m $ corresponding to a potential $V_0 = -E_0 l = - 4 \pi \sigma_m l$. In the presence of dielectric material the applied electric field $E_0$ is screened by the induced polarization $P_x$.   The resulting macroscopic electric field $E$ can be written as
\begin{equation}
\label{excap}
 E = 4 \pi \left( \sigma_m + \sigma_p \right) = E_0 - 4 \pi P_x
\end{equation}  
where $\sigma_p = - P_x$ is the polarization charge density accumulating on the  surface of the dielectric (see Fig.~\ref{fig0}). Because the polarization is aligned along the applied field, $E_0$ and $P_x$ have the same sign (positive in the figure). $\sigma_p$ and $\sigma_m$ have opposite sign. The macroscopic electric field is determined by the net interface charge $\sigma_m+ \sigma_p$ and therefore $E < E_0$. Similarly the potential $V= - E l$ is lower (in absolute value) than $V_0$. This is how capacitors store charge\cite{Purcell:2011ca}.    

 Setting $E_x=E, E_y=E_z=0$ in Eq.~\ref{fhmd} we obtain the electric enthalpy hamiltonian
\begin{equation}    
\label{fhmdx}
\mathcal{F}_x\left( E, v \right) = H_{\mathrm{PBC}}(v)- \mathit{\Omega} E P_x(v)
\end{equation}
How can  this relatively  simple MD model without interfaces possibly represent the capacitor of Fig.~\ref{fig0}?  The idea is that interactions not affected by surface effects are described by the supercell hamiltonian $H_{\mathrm{PBC}}$. For a macroscopic capacitor these interactions are assumed to include  everything except the coupling to the electric field which is accounted for in the  $-\mathbf{E} \cdot \mathbf{P} = EP_x$ term. This term plays the role of a pair of ``virtual'' electrodes connected to a voltage source imposing a potential drop of  $\Delta V = - E L$ over the length $L$ of the MD cell in the $x$ direction. $ \mathcal{F}_x\left( E, v \right)$ is a microscopic Hamiltonian, not a free energy. All quantities, except $E_x= E$ fluctuate in time. This applies to the polarization charge $\sigma_p(t) = - P_x(v(t))$ but also to the charge on the virtual electrode $\sigma_m(t)$. It is only the sum $\sigma_m(t) + \sigma_p(t) = E/4 \pi$ that is constant. The instantaneous electrode charge compensating the varying polarization charge is supplied by the voltage source. The implication is that for  $E=0$  Eq.~\ref{fhmdx} can be viewed as a capacitor in short circuit, consistent with the accepted view of the Ewald summation method.

Next we introduce the displacement field. For the parallel plate capacitor $D_x = E_0 = 4 \pi \sigma_m$. As explained above, $D_x$ is interpreted as a property of the microscopic system of Eq.~\ref{fhmdx}  and fluctuates in time  because the electric field $E_x = E$ is fixed. Applying the formalism of section \ref{sec:EDham}, $D_x$, in turn, can be constrained to a value $D$ in the dynamics driven by the Hamiltonian  
 \begin{equation}
\label{uvdbx}
\mathcal{U}_x\left(D, v \right) = H_{\mathrm{PBC}}(v)+ \frac{\mathit{\Omega}}{8 \pi} \left( D - 4 \pi P_x(v) \right)^2
\end{equation}
The constraint on the displacement field applies only to the $x$ direction. Eq.~\ref{uvdbx} is not a special case of Eq.~\ref{uvdb}. In the $y$ and $z$ directions the regular Ewald boundary conditions are maintained and hence $E_y =0$ and $E_z=0$.   Because  $D$ is effectively the charge density on the virtual electrodes ($D = 4 \pi \sigma_m$),  fixing $D$ in Eq.~\ref{uvdbx}  corresponds to simulating an  open circuit capacitor. Now it is the conjugate variable $E_x(t)$ that fluctuates. The time dependence of $E_x$ is passed on to the potential $\Delta V = - E_x L$ across the cell.  

As pointed by SSV Eq.~\ref{uvdbx} can be regarded as a hybrid form of Eq.\ref{uvdb}, obtained by a partial Legendre transform of the Hamiltonian of Eq.~\ref{fhmd} (see further Ref.~\citenum{Stengel:2009cd}).  In a plate capacitor $\mathbf{D} = \mathbf{E}_0$ on average.  However, it would be wrong to identify  $\mathbf{D}(t)$ with $\mathbf{E}_0$ at every instant of time. The applied field, in the orientation of Fig.~\ref{fig0}, is strictly along the $x$ axis. While on average the $y$ and $z$ components of $\mathbf{D}$ vanish, instantaneous values can be finite.  $D_y(t)$ and $D_z(t)$ are equal to the transverse polarization  $P_y(t)$ and $P_z(t)$\cite{Kivelson:1989fj}. In fact, compared to the longitudinal polarization  transverse polarization ($P_x(t)$ in our model capacitor) shows substantially larger fluctuations  (see section \ref{sec:fluc}). 

\section{Dielectric constant} \label{sec:diel}

\subsection{Dielectric constant from polarization fluctuations} \label{sec:fluc}
The constant $\mathbf{E}$  and  $\mathbf{D}$  ensembles introduced in section \ref{sec:ED}  apply to different electrical boundary conditions.  Fluctuations of the polarization $\mathbf{P}$ are therefore expected to differ in magnitude  and may occur on different timescales.  However, following Kirkwood-Fr\"ohlich theory~\cite{Kirkwood:1939br, frohlich1958theory}, it should be possible to obtain an estimate of the dielectric constant from polarization fluctuations either  under constant $\mathbf{E}$  or  $\mathbf{D}$ dynamics.  Staying with the capacitor paradigm of section \ref{sec:cap} application of linear response to the system defined by the electrical enthalpy hamiltonian Eq~\ref{fhmdx} gives
\begin{equation}
\label{chi}
\langle P_x \rangle= \beta \mathit{\Omega} \left( \langle P_x^2\rangle -
  \langle P_x\rangle^2 \right) E
\end{equation}
where the second moment is evaluated at zero field ($E=0$).  The liquid is isotropic and we can set 
$\langle P_z^2\rangle -  \langle P_z\rangle^2=(\langle \mathbf{P}^2\rangle -
  \langle \mathbf{P}\rangle^2)/3$, which gains us some accuracy in the statistics.   The response coefficient of Eq.~\ref{chi} is the susceptibility $\chi$. Converting to the dielectric constant using $4 \pi \chi = \epsilon - 1 $ we find
\begin{equation}
\label{epse}
\epsilon =1+\frac{4\pi\beta \mathit{\Omega}}{3} \left(\langle \mathbf{P}^2\rangle_{\mathbf{E}=0} -
  \langle \mathbf{P}\rangle^2_{\mathbf{E}=0} \right)
\end{equation}
For clarity the thermal average brackets in Eq.~\ref{chi} have been marked with a subscript indicating  the condition under which the fluctuations have been obtained. 

Eq.~\ref{epse}  is identical to the standard fluctuation formula used in Ewald summation\cite{DeLeeuw:1980rs80a,DeLeeuw:1980rs80b,Neumann:1983kg,Neumann:1984mp}. We arrived at this established result without having to worry about how to relate the macroscopic field  $\mathbf{E}$ to the applied field $\mathbf{E}_0$\cite{Neumann:1983gz,Neumann:1983kg,Madden:1984cp}. The electric field in the SSV extended Hamiltonian is directly equal to the macroscopic field. This is a key feature of the SSV scheme. We will return to this important point once more in  section \ref{sec:physchem} where we make a comparison to the methods  commonly used in computational physical chemistry.

Alternatively we can average $P_x$ over the ensemble defined by  the  open circuit Hamiltonian of Eq.~\ref{uvdbx}.   It is not difficult to show that linear response now leads to the relation
 \begin{equation}
\label{alpha}
\langle P_x \rangle= \beta \mathit{\Omega} \left( \langle P_x^2\rangle -
  \langle P_x\rangle^2 \right) D
\end{equation}
In Eq.~\ref{alpha} we recognize the definition of the polarizability $\alpha$.  Eq.~\ref{alpha} can be exploited to obtain another estimate of the dielectric constant via the relation  $ 4 \pi\alpha = 1 - 1/\epsilon$  which should be consistent with the estimate from the susceptibility $\chi$  (Eq.~\ref{chi}).  

Similar to Eq.~\ref{epse} we would like to retain the extra boost in accuracy provided by isotropy (which is even more critical here, see section \ref{sec:results}). However, there is a complication. Eq.~\ref{alpha} resembles Eq.~\ref{chi}, but in contrast to  a short circuit capacitor, the open circuit system is not isotropic (see the discussion in  in section \ref{sec:cap}).  Fluctuations in the $y$ and $z$ (transverse) direction are distinct from the longitudinal fluctuations in $P_x$.  Isotropy can be restored by imposing a $D=0$ constraint also in the $y$ and $z$ direction which amounts to using the hamiltonian of Eq.~\ref{uvdb} for $\mathbf{D}=0$.   The corresponding estimate for $\epsilon$  obtained from the polarization fluctuations is written as
\begin{equation}
\label{epsd}
\epsilon =\frac{1}{1-4\pi\beta \mathit{\Omega} (\langle \mathbf{P}^2\rangle_{\mathbf{D}=0} -
  \langle \mathbf{P}\rangle^2_{\mathbf{D}=0})/3}
\end{equation}
However, it is not immediately clear what $\mathbf{D}=0$  ensemble represents. This is indeed an important question for the understanding of the SSV method and we will come back to it  in section \ref{sec:physchem} and appendices \ref{sec:surf} and \ref{sec:reac}.

Finally, rearranging Eq.~\ref{epse} and Eq.~\ref{epsd} leads to:
\begin{equation}
\label{ratio}
\epsilon =\frac{\langle \mathbf{P}^2\rangle_{\mathbf{E}= 0} -
  \langle \mathbf{P}\rangle^2_{\mathbf{E} = 0}}{\langle
  \mathbf{P}^2\rangle_{\mathbf{D}=0} -  \langle \mathbf{P}\rangle^2_{\mathbf{D}=0}}
\end{equation}
As shown in the Madden and Kivelson's review~\cite{Madden:1984cp}(see also Ref.~\citenum{Kivelson:1989fj}) polarization fluctuations  are anisotropic in the  $\mathbf{k} \rightarrow 0$ limit, even if the dielectric tensor is isotropic.  Transverse and longitudinal fluctuations  differ  by a factor  $\epsilon$ (for non-polarizable polar molecules).  The same ratio is found in Eq.~\ref{ratio} supporting our hypothesis (our argument is not a  proof) that  the fluctuations sampled at constant $\mathbf{E}$ and constant $\mathbf{D}$ can be identified with  the $\mathbf{k}=0$ limit of transverse respectively  longitudinal polarization. We furthermore note that a relation similar to Eq.~\ref{ratio} can be derived in the framework of reaction field methods\cite{Neumann:1983gy,Neumann:1983kg}. This is discussed in Appendix \ref{sec:reac}.

\subsection{Dielectric constant from finite field derivatives} \label{sec:finfield}

Simulations at finite field should enable us to determine $\epsilon$ directly from the field derivative of polarization, which we then can compare to the fluctuation estimates at zero field.  For sufficiently small fields these results should in principle agree but will differ in practice because of different requirements on the accuracy of the sampling and possible finite system size effects. An assessment of these effects   is the main purpose of our investigation.  Finite field calculations are also of interest as a test of the SSV scheme for additional reasons.  While fluctuations  vary with  the electrical boundary conditions, the expectation value of $\mathbf{P}$ is a thermodynamic state variable which should, in the thermodynamic limit, be the same for a given thermodynamic state, whether obtained under constant $\mathbf{E}$ or constant $\mathbf{D}$.
 
For a  sufficiently small electric field the dielectric constant can be estimated using the relation
\begin{equation}
\label{epse2}
\epsilon=1+\frac{4\pi\langle P_x \rangle}{E}
\end{equation}
with $\langle P_x \rangle$ the expectation value of polarization obtained from a MD run using the  constant $E_x =E$ Hamiltonian of Eq.~\ref{fhmdx}. A similar equation, valid in the linear regime, was already used to extrapolate to zero-field  by Yeh and Berkowitz in their finite field calculation of $\epsilon$ of liquid water~\cite{Yeh:1999hb}.  The corresponding finite $D$ estimate for $\epsilon$  follows from an  inverse relation
\begin{equation}
\label{epsd2}
\epsilon=\frac{1}{1-4\pi\langle P_x \rangle/D}
\end{equation}
where the expectation value of polarization is determined from an average over a trajectory generated by the constant $D_x = E_0 =D$ Hamiltonian of Eq.~\ref{uvdbx}.

\subsection{Constant applied electric field $\mathbf{E}_0$} \label{sec:physchem}
The form of the SSV electric enthalpy hamiltonian of Eq.~\ref{fhmd}  might be, at first, a surprise for readers familiar with the physical chemistry literature on polar liquids expecting to see the applied electric field $\mathbf{E}_0$ in the coupling term (see for example Refs.~\citenum{Neumann:1983kg,Neumann:1984mp} and \citenum{Madden:1984cp}). Eq.~\ref{fhmd} is however a microscopic electric enthalpy intended for constant macroscopic field $\mathbf{E}$, not constant applied electric field $\mathbf{E}_0$. The difference $ \mathbf{E}_p = \mathbf{E} -  \mathbf{E}_0$ is the polarization field, also referred to as the depolarising field in literature. $\mathbf{E}_p$ is the electric field generated by the polarization. Substituting $\mathbf{E} = \mathbf{E}_0 +  \mathbf{E}_p$ in Eq.~\ref{fhmd} gives
 \begin{equation}
\label{fhmd0}
\mathcal{F}\left( \mathbf{E}, v \right) = H_{\mathrm{PBC}}(v)- \mathit{\Omega}  \left( \mathbf{E}_0(v)  + \mathbf{E}_p (v) \right) \cdot \mathbf{P}(v)
 \end{equation}
Similar to the polarization, $\mathbf{E}_0$ in Eq.~\ref{fhmd0} depends on the microstate $v$. It is not constant all. 

The parallel plate capacitor of section \ref{sec:cap} is again the best example to understand Eq.~\ref{fhmd0}. In this geometry $ E_p = - 4 \pi P_x = 4 \pi \sigma_p $ (see Eq.~\ref{excap}). The depolarizing field is determined by the polarization surface charge $\sigma_p$, while the applied field $E_0= 4 \pi \sigma_m$  is generated by the electrode charge $\sigma_m$. To keep the voltage constant, the voltage source adds or removes electrode charge compensating for the fluctuation in the polarization charge.  Formulated in terms of fields, the applied electric field responds instantaneously to the fluctuations in the depolarizing field such that the sum $E = E_0 - 4 \pi P_x$ = constant.  Substituting in the enthalpy Hamiltonian for the capacitor (Eq.~\ref{fhmdx}) we obtain
\begin{equation}    
\label{fhmdx0}
\mathcal{F}_x\left( E, v \right) = H_{\mathrm{PBC}}(v)- \mathit{\Omega} (E_0(v) - 4 \pi P_x(v)) P_x(v)
\end{equation}
Even for a shortcircuited capacitor ($V=E=0$),  the applied field, while zero on average, will have finite instantaneous values, cancelling the fluctuations in $E_p$.  To control $E_0$ we must disconnect the voltage source (open circuit). The charge on the metal electrode is now fixed and therefore $E_0$. The Hamiltonian to use for constant $E_0$ is therefore not Eq.~\ref{fhmdx} but Eq.~\ref{uvdbx}. 

Can Eq.~\ref{uvdbx} be rewritten in a form more recognizable to physical chemists? To answer this question we expand the coupling term (omitting volume for simplicity)     
 \begin{equation}
\label{uvdbx0}
  \frac{1}{8 \pi} \left( D - 4 \pi P_x(v) \right)^2 =
 \frac{E_0^2}{8 \pi} - E_0 P_x(v) + 2 \pi P_x(v)^2 
\end{equation}
where we have used that for the parallel plate capacitor $D=E_0$. The first term of Eq.~\ref{uvdbx0} gives the energy of the polarizing field. The second term is the sought for coupling of the polarization to the external field. To identify the last term we introduce the depolarizing field $E_p = -4 \pi P_x$ to find that the third term is equal to $ - E_p P_x/2$, the work done against the depolarizing field in the process of polarization. This energy has been shown to the equal to the surface term in Ewald summation carried out over layers\cite{Smith:1981rsa,Ballenegger:2014jpc} instead of the more popular summation over spherical shells\cite{DeLeeuw:1980rs80a}. The key point here is that the interaction of polarization with its own electric field is not included in the Ewald Hamiltonian $H_{\mathrm{PBC}}$. The coupling to the depolarizing field must be accounted for explicitly as part of the extended hamiltonian.  In conclusion  Eq.~\ref{uvdbx} seems indeed equivalent to the hamiltonians used in theory of a polar liquids\cite{Madden:1984cp,Kivelson:1989fj} if we assume that the interaction is implicit in the Hamiltonian.

There are further similarities linking Eq.~\ref{uvdbx0} to existing methods. The surface term $2 \pi P_x^2$ is well-known in the literature on simulation of liquid-solid interfaces\cite{Yeh:1999dm,Ballenegger:2005jpc,Bonthuis:2012lm}.  A benchmark in the field is the 1999 paper by Yeh-Berkowitz  who introduced a correction term to decouple the electrostatic interactions between a slab of material and its periodic images\cite{Yeh:1999dm}. The same term is used in surface science known there as the dipole correction\cite{Neugebauer:1992uh,Bengtsson:1999il}. These two corrections are identical and are moreover equal to the coupling term in Eq.~\ref{uvdbx0} for $E_0=0$.  Note, however, that interface/surface modelling and the SSV scheme aim for different target systems.  The ideal model system in computational surface science is an isolated slab suspended in vacuum. A similar setup is often used to model liquid-solid interfaces inserting a vacuum spacer in the solid. Periodic models include therefore vacuum layers of a width comparable to or larger than the particle system. The Yeh-Berkowitz correction is intended as a computationally efficient replacement of the costly 2D Ewald sum method (it seems to be doing a very good job as shown in Ref.~\citenum{Yeh:1999dm}).  SSV models, on the other hand, are designed to  represent continuous condensed phase systems. These systems can be homogeneous such as liquid water studied here.  Vacuum layers opening up unwanted interfaces are avoided.

Finally, moving on to the general SSV constant $\mathbf{D}$ Hamiltonian of section \ref{sec:EDham} we set $D_x = D, D_y = D_z = 0$ in Eq.~\ref{uvdb} to obtain
  \begin{eqnarray}
\label{uvdbxyz}
  \mathcal{U}(D,v) & = & H_{\mathrm{PBC}}(v) + \frac{\mathit{\Omega}}{8 \pi}
 \left(D-4\pi P_x(v)\right)^2 \\ \nonumber
  & & + 2 \pi \mathit{\Omega} \left(P_y(v)^2 +  P_z(v)^2 \right)  
\end{eqnarray}
The coupling term is different from the one in Eq.~\ref{uvdbx0} for the open circuit parallel plate capacitor. In Eq.~\ref{uvdbxyz} adds further quadratic terms for the polarization in the perpendicular $y$ and $z$ direction. However, while Eq.~\ref{uvdbxyz} may look unfamiliar, or even unphysical from the perspective of physical chemistry, it is supported by a thermodynamic foundation via Eq.~\ref{ework} giving it a special status among other forms of constant applied field hamiltonians\cite{Vega:2011prl,Kolafa:2014jctc,vanGunsteren:2011jctc}.

Setting $D=0$ in Eq.~\ref{uvdbxyz} we obtain
\begin{equation}
\label{u0vdb}
  \mathcal{U}_{D=0}(v)  =  H_{\mathrm{PBC}}(v) + 2 \pi \mathit{\Omega} \mathbf{P}^2
\end{equation}
This Hamiltonian is of special interest as it used to sample the $\mathbf{D}=0$ fluctuations in Eq.~\ref{epsd}. The  self interaction term in Eq.~\ref{u0vdb} resembles the $(2 \pi \mathit{\Omega}/3) \mathbf{P}^2$ surface term in Ewald summation for spherical vacuum boundary conditions\cite{DeLeeuw:1980rs80a} but is however a factor three larger. The difference can be traced back to the finite transverse polarization of a polarized sphere surrounded by vacuum.  Further discussion is deferred to appendix \ref{sec:surf}. Surprisingly, the Hamiltonian Eq.~\ref{u0vdb} is known in the framework of the reaction field method\cite{Neumann:1983gy,Neumann:1983gz,Neumann:1983kg}. Eq.~\ref{u0vdb} can be reproduced by setting the dielectric constant of the embedding dielectric continuum to zero\cite{Caillol:1994jcp,Maggs:2004jcp}(for details see Appendix \ref{sec:reac}). Such an Hamiltonian has in fact been used for simulation studies of the dielectric properties of ionic solutions\cite{Caillol:1989jcp}. The Hamiltonian, while useful, was regarded as somewhat unphysical. It is not in the context of SSV theory. Moreover, Anthony Maggs has pointed out that an Hamiltonian of the form Eq.~\ref{u0vdb} can also be obtained from the time dependent Maxwell equations\cite{Maggs:2004jcp,Maggs:2006prl}). His argument is summarized in Appendix \ref{sec:surf}.

\subsection{Model system and molecular dynamics} \label{sec:md}
The theory outlined above was verified by a classical MD simulation of liquid water at ambient conditions. The system consists of 706 water molecules in a fixed cubic box with length 27.7~\AA. The interactions are described by the SPC/E water model~\cite{Berendsen:1987uu}. The molecules are kept rigid using the SETTLE algorithm~\cite{Miyamoto1992}. The MD integration time step is 2~fs. The Ewald summation is implemented using the Particle Mesh Ewald (PME) scheme\cite{Ewald}. Short-range cutoffs for the van der Waals and Coulomb interaction in the direct space are 10~\AA. The temperature is controlled by a Nos\'e-Hoover chain thermostat~\cite{martyna92} set at 298K. All simulations are done with a modified version of GROMACS 4 package~\cite{Hess2008}. 

Two points need further comments on the practical implementation in classical force field based MD: one is the computation of the macroscopic polarization and the other is the force calculation in constant $\mathbf{E}$ and constant $\mathbf{D}$ simulations. Polar liquids such as water are extended systems. The total dipole moment of a periodic supercell depends, in principle, on how we decide to draw the boundaries~\cite{King-Smith:1993prb,Resta:1994rmp,Resta:2007ch} . If the bonds of a molecule are cut by the boundaries, the two halves of this molecule will end up on opposite sides of the MD cell resulting in a huge change of the dipole moment. This problem can be ignored in practice because the molecular structure provides a natural gauge and it is automatically done for the rigid water model used in simulations. Therefore, the macroscopic polarization can be simply defined as the sum of the dipole moments of the molecules. For constant $\mathbf{E}$ simulation, the field-dependent force on atom $i$ is $q_i\mathbf{E}$, where $q_i=\frac{\partial \mathbf{P}}{\partial\mathbf{r}_i}$ is the point charge assigned to the atom in the SPC/E model.  For constant $\mathbf{D}$ simulation, the field-dependent force on atom $i$ is $q_i\mathbf{D} - 4\pi q_i\mathbf{P}$. In this case, the force depends explicitly on the value and continuity of the macroscopic polarization.

\section{Results} \label{sec:results}
 
\subsection{Structure and dynamics} \label{mdres}
\begin{figure}
\includegraphics[width=0.9\columnwidth]{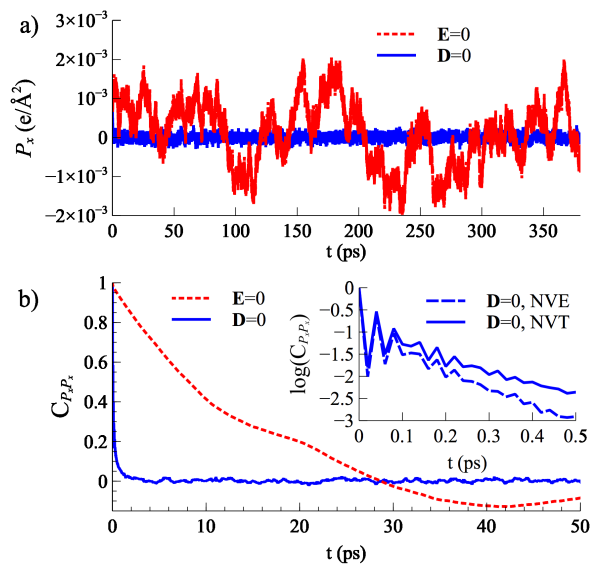}
\caption{\label{fig1} Simulation of bulk liquid water at $\mathbf{E}=0$ using the hamiltonian of Eq.~\ref{fhmd} and $\mathbf{D}=0$ using the hamiltonian of Eq.~\ref{uvdb}: a) Time evolution of $P_x$, the $x$ component of the polarization; b) Corresponding autocorrelation function defined as $C_{P_xP_x}= \langle
 P_x(0)P_x(t) \rangle/\langle P_x(0)P_x(0)\rangle$. The inset shows the short time behaviour of  $C_{P_xP_x}$ for $\mathbf{D} = 0$.}
\end{figure}

Theory predicts (Eq.~\ref{ratio}) that a $\mathbf{D}=0$ constraint has the effect of suppressing polarization fluctuations compared  to $\mathbf{E}=0$ conditions.  The corresponding relaxation times are also faster. This is shown in  Fig.~\ref{fig1} for the $x$ component (the system is isotropic, so $P_y$ and $P_z$ behave the same).  The mean of $P_x$ vanishes for both the  $\mathbf{E}=0$  and $\mathbf{D}=0$ time series (Fig.~\ref{fig1}a)) but the amplitude of the $\mathbf{E}=0$  oscillations is significantly larger.  

For an analysis of the time dependence  (Fig.\ref{fig1}b) it is useful to recall the classical Debye theory of the relaxation of polarization. The relaxation  in Debye theory is exponential.  Indeed, as Fig.~\ref{fig1}b shows, the autocorrelation function of $P_x$ at $\mathbf{E}=0$ decays exponentially (with a hint of a slow oscillation). The relaxation time is 10.3~ps, which is close to the experimental Debye relaxation time $\tau_D$  for water.  $\tau_D$ is the time for response to a sudden change in the electric field $\mathbf{E}$.  The other relaxation time defined in Debye theory is $\tau_L$ controlling the response to a change in $\mathbf{D}$ (or equivalently a jump in the charge of a solute).   $\tau_L = \tau_D/\epsilon$ and $\tau_L$ and $\tau_D$ can be interpreted as a longitudinal respectively transverse relaxation time\cite{Madden:1984cp,Kivelson:1989fj}. In agreement with this picture,  we find that switching from $\mathbf{E}=0$ to  $\mathbf{D} =0$  accelerates the relaxation, but the time dependence in the open circuit system appears to more complex. The short time behaviour (see inset) clearly shows oscillations. Estimating an effective decay time from the time envelope, we obtain $\tau_L= 0.3$ ps. 
 
\begin{figure}
\includegraphics[width=0.85\columnwidth]{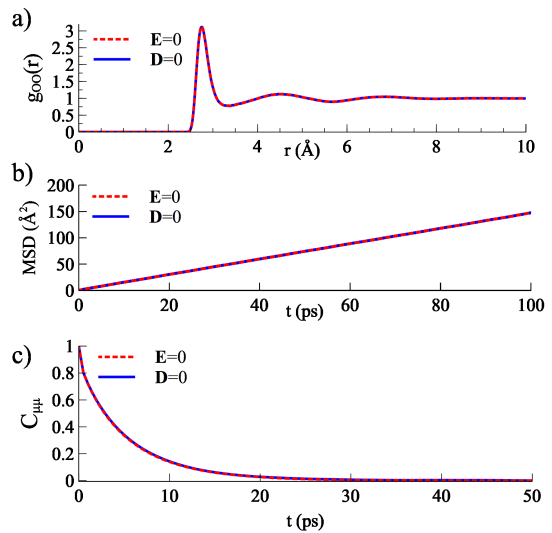}
\caption{\label{fig2} Simulations of bulk liquid water at
  $\mathbf{E}=0$ and $\mathbf{D}=0$. a) The oxygen-oxygen radial distribution
  functions $g_{\text{OO}}(r)$; b) The mean squared displacements (MSD)
  of water molecules; c) Autocorrelation function of the molecular dipole moment $C_{\mu\mu}$. }
\end{figure}

The pronounced contrast in magnitude and time scale of polarization fluctuations raises the question whether this collective behaviour is reflected in the local molecular structure and dynamics. Fig.~\ref{fig2} shows the oxygen radial distribution function, molecular diffusion rate as measured by the mean square displacements and autocorrelation function of the molecular dipole moment characterizing molecular orientation.  These properties are often used as probe of the structure and dynamics in liquid water at the single molecular level. As can be seen from Fig.~\ref{fig2} the change in electrical boundary condition has little or no effect on the translational and orientational motion of the water molecules.

\subsection{Static dielectric constant from fluctuations} \label{sec:gk}
The formalism of sections \ref{sec:fluc} and \ref{sec:finfield} gives us four different estimates of $\epsilon$, two from fluctuations, Eq.~\ref{epse} and Eq.~\ref{epsd}, and two from finite field derivatives, Eq.~\ref{epse2} and \ref{epsd2}. We start with the fluctuation approach. 

\begin{figure}
\includegraphics[width=0.9\columnwidth]{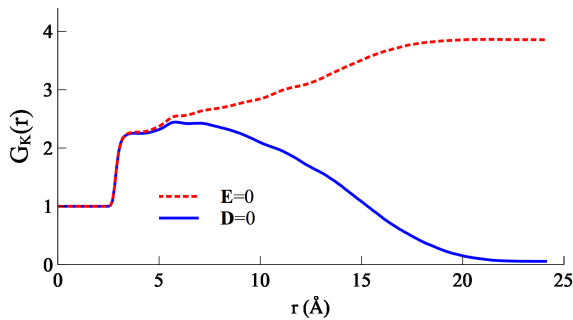}
\caption{\label{fig3} Comparison of the distance dependence of the Kirkwood G-factor $G_{\mathrm{K}}(r)$ evaluated under $\mathbf{E}=0$ and $\mathbf{D}=0$ constraints}
\end{figure}

The estimates of $\epsilon$ for SPC/E calculated from Eq.~\ref{epse} and Eq.~\ref{epsd} are 71.4(8) and 76(4) in good mutual agreement and with literature values ranging from 67 to 81~\cite{vanderSpoel:1998gb, Aragones:2011dp, Zhang:2014jq, Braun:2014fs}. While the dielectric constant estimate should be the same, whether determined under $\mathbf{E}=0$ or $\mathbf{D}=0$ constraints, the polarization fluctuations under these conditions are very different. This is reflected in the $r$ dependent Kirkwood G-factor $G_{\mathrm{K}}(r)$, which is a orientational correlation function for (rigid) dipoles(see for example Ref.~\citenum{Neumann:1983kg}). It is defined as   
\begin{equation} 
\label{GK}
G_{\mathrm{K}}(r)=1+ N(r) \sum_{j,\,r_{ij}<r} \langle \cos\theta_{ij}\rangle
  \end{equation}
The sum is over all molecules $j$ enclosed in a sphere of radius $r$ centered on molecule $i$. $N(r)$ is the number of molecules in the sphere. $\theta_{ij}$  is the angle between the dipole $\boldsymbol{\mu}_j$ of molecule $j$  and the dipole $\boldsymbol{\mu}_i$ of the central molecule $i$. Our results for $G_{\mathrm{K}}(r)$ are plotted in Fig.~\ref{fig3}. Both $\mathbf{E}=0$ and $\mathbf{D}=0$ curves settle in a radius independent asymptotic regime for distance $r > 22$\AA. The Kirkwood G-factor interpolates between local and global behaviour. The variation with $r$ at short range ($r<6$\AA) is similar for  $\mathbf{E}=0$ and $\mathbf{D}=0$. The two curves part for increasing values of $r$. 

\begin{figure}
\includegraphics[width=0.9\columnwidth]{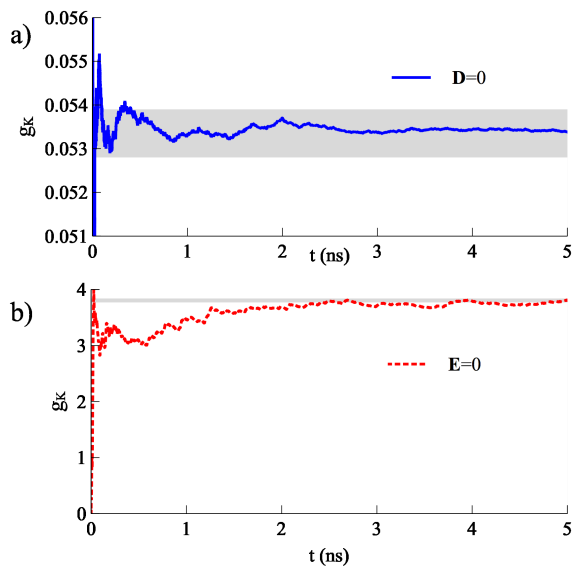}
\caption{\label{fig4} The accumulating average of the normalized variance of the total dipole moment $g_K$ of the MD cell with the length of the MD run. $g_K= (\langle \mathbf{M}^2 \rangle-\langle \mathbf{M}\rangle^2)/ (N\mu^2) $, where $N$ is the number of water molecules, $\mu$ is the (fixed) dipole moment of a single water molecule and $\mathbf{M}= \sum^N \boldsymbol{\mu}_i$.  The shaded area is the margin for a 1\% deviation from the final average.}
\end{figure}

Calculations of the static dielectric constant cannot be presented without error analysis. As demonstrated in Refs.~\citenum{Elton:2014jcp} and \citenum{vanGunsteren:2011jctc} finite size effects are less of a concern for system sizes accessible to classical MD simulation. The MD cell used here containing 706 water molecules should be large enough for the purpose. The time scale needed to converge a second moment of the total dipole moment is a more critical issue. This is confirmed by the accumulating average of the normalized variance of the total dipole moment $g_K$ determined with regular Ewald summation ($\mathbf{E}=0$) shown in Fig.~\ref{fig4}a. Consistent with the literature we find that it takes at least several nanoseconds to reduce the statistical uncertainty to a value below 1\%. As can be seen from Fig.~\ref{fig4}b, the same accuracy is reached within less than one nanosecond by changing the electrical boundary conditions to $\mathbf{D}=0$. Unfortunately, because of the troublesome inverse relation between fluctuations and dielectric constant (Eq.~\ref{epsd}) the accuracy in the second moment must be proportionally higher, and much of the apparent gain in time scale is lost in practice.   

Returning to the question discussed in section \ref{sec:physchem} how to compare the SSV method to methods used in physical chemistry, we note that the sensitivity of the Kirkwood G factor to a change of boundary condition has been studied in detail by Neumann for the Stockmayer fluid\cite{Neumann:1983kg}. Neumann builds on the familiar cavity model of Kirkwood. His systems consist of a sphere containing the Stockmayer atoms (point dipoles with short-range Lennard-Jones pair interactions) embedded in a dielectric continuum. The dielectric constant of the continuum $\epsilon^\prime$ is varied from $\epsilon^\prime=\infty$ (conducting) to $\epsilon^\prime=1$ (vacuum). The distance dependent Kirkwood factors Neumann obtains for conducting and vacuum boundary conditions have a clear resemblance to our results of Fig.~\ref{fig3} for liquid water with $\mathbf{E}=0$ corresponding to $\epsilon^\prime = \infty$ and $\mathbf{D} =0 $ to $\epsilon^\prime=1$. For conducting boundary conditions this was expected because, as mentioned, the expression relating the dielectric constant to the dipole fluctuations (Eq.~\ref{epse}) agree. However, the expression derived by Neumann for the ratio of the $\epsilon^\prime=\infty$ and $\epsilon^\prime=1$ total dipole  fluctuations is $(\epsilon+2)/3$, which (for large $\epsilon$) is a factor three smaller than what we obtained in Eq.~\ref{ratio}. Indeed, the ratio between the normalized variance of the total dipole moment $g_K$ at $\mathbf{E}=0$ and $\mathbf{D}=0$ from our simulations gives 71 directly validating Eq.~\ref{ratio}. This again raises the question about a possible geometric interpretation of the SSV $\mathbf{D}=0$ Hamiltonian. This will be discussed in detail in appendix \ref{sec:surf}.           

\subsection{Dielectric constant from field derivatives} \label{sec:mdfield}
Finite $E$ simulations necessarily involve a limited subset of state points. We selected five $E_x = E$ values with increasing strengths, as listed in the first column of Table~\ref{tab1}. For these five values we carried out constant $E_x$ simulations using the Hamiltonian of Eq.~\ref{fhmdx} determining for each of these runs the average of $P_x$ which is indicated in the Table.  Next we used  Eq.~\ref{ddef} to compute the $D$ values corresponding to the $P_x$ we had obtained. These values of $D$ were then taking as the displacement field in  constant $D_x$ simulation using the Hamiltonian of Eq.~\ref{uvdbx}.  If the SSV constant $\mathbf{D}$ method works as promised, the resulting $P_x$ are the same as those of the constant $E_x$ simulations from which the $P_x$ values were sampled.  Indeed, as shown in the second and fourth columns of Table~\ref{tab1}, this is the case. Note that the values of $D$ are an order of magnitude larger compared to $E$ for the same value of $P_x$, reflecting the efficient dielectric screening in liquid water. 

\begin{table}
\caption{\label{tab1} Simulation conditions at  constant $E_x$
 or constant $D_x$ and the corresponding observed $\langle P_x \rangle$.}
\begin{ruledtabular}
\begin{tabular}{ l l l l }
$E_{x}$ (V/\AA) & $\langle P_x \rangle$ ($10^{-3}$ e/\AA$^2$) &
$D_x$ (V/\AA) & $\langle P_x \rangle $  ($10^{-3}$ e/\AA$^2$) \\
\hline
0.01 &3.72(3) &0.684&3.724(2)\\
0.02 & 6.41(5) & 1.180& 6.418(2) \\
0.04 &9.66(6) &1.788  &9.660(1) \\
0.10 &12.62(3) &2.385&12.632(1) \\
0.28 &14.507(4) &2.907 &14.512(1) \\ 
\end{tabular}
\end{ruledtabular}
\end{table}

\begin{figure}
\includegraphics[width=0.9\columnwidth]{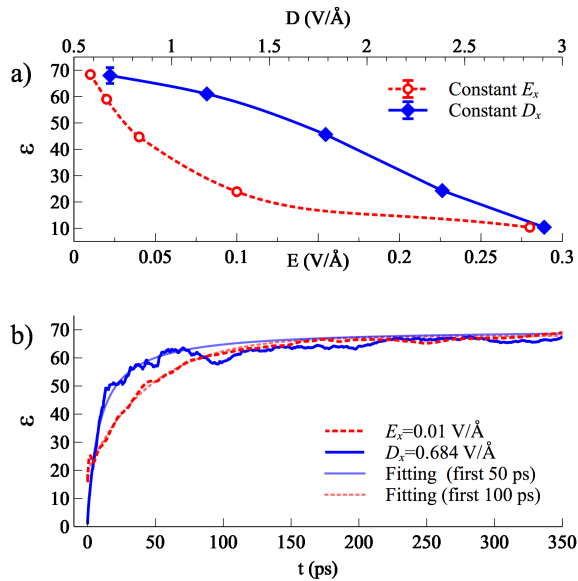}
\caption{\label{fig5} a) The static dielectric constant $\epsilon$ at
 constant $E$ and constant $D$; b) The accumulating average of
 $\epsilon$  at $E_z=0.01$ V/\AA~and $D_z=0.684$ V/\AA. }
\end{figure}

After this crucial consistency test, we computed the $E$ and $D$ derivative estimate of the dielectric constants using the method explained in section \ref{sec:finfield}. The comparison of $\epsilon$ obtained form the polarizability as a function of $D$(Eq.~\ref{epsd2}) to $\epsilon$ computed from the susceptibility as a function of $E$(Eq.~\ref{epse2}) is plotted in Fig.~\ref{fig5}a. Regarding statistics and convergence, the small values of the field are the most critical and computer time consuming. Fig.~\ref{fig5}b gives the running average for the state point corresponding to our smallest electric field. Comparison to Fig.~\ref{fig4} confirms that the convergence for averages of the polarization is still faster than for the second moment even for small fields. This is of course as expected. It is encouraging to see that the convergence time of the dielectric constant under constant $D_x$ turns out to be shorter than for constant $E_x$, even with the unfavourable inverse relation of Eq.~\ref{epsd2} where the relative error $\delta \epsilon/\epsilon$ is proportional to $\epsilon$.

\begin{figure}
\includegraphics[width=0.9\columnwidth]{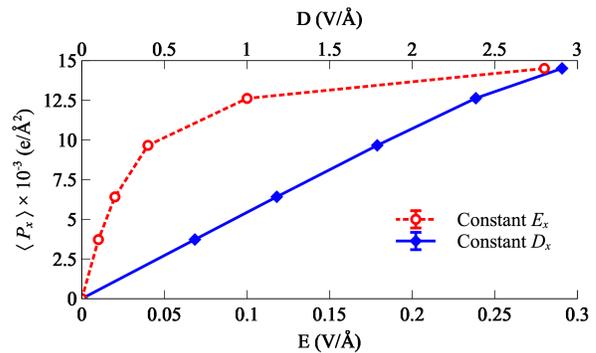}
\caption{\label{fig6} Polarization as a function of electric field $E$ and displacement $D$ determined from constant $E$ respectively constant $D$ molecular dynamics.}
\end{figure}

Using Eqs.~\ref{epse2} and \ref{epsd2} the way we did in Fig.~\ref{fig5} amounts to a global linear approximation. If the dielectric response of water was linear in the range of fields investigated the curves in Fig.~\ref{fig5}a would be horizontal straight lines. Not surprisingly, they are not. The variation of $\epsilon$ with field strength is not even linear. Dielectric saturation for increasing $E$ is known to follow the so-called Debye-Langevin equation ($\epsilon \sim 1/E(\coth(\beta\mu E)-1/\beta\mu E)$), derived for an independent dipole approximation~\cite{debye1929polar, Booth:1951ga}. Consistent with this simple picture the $\epsilon(E)$ dependence obtained from the constant $E$ simulations is approximately exponential approaching an asymptotic value at about $E= 0.2$V/\AA{}. The curvature in $\epsilon(D)$ is opposite to the curvature in $\epsilon(E)$. The non-linear effect in the $\epsilon(D)$ curve is much less pronounced. This is also evident from a direct comparison of the $P_x(D)$ to the  $P_x(E)$ dependence(Fig.~\ref{fig6}). This effect is, from technical point of view, perhaps the most encouraging observation made in this study, because it makes the extrapolation to zero field easier. 

\section{Summary and outlook} \label{sec:concl}

Including a term coupling polarization to electric fields is the first step in the derivation of fluctuation expressions for dielectric response coefficients of polar liquids in Kirkwood-Fr\"ohlich theory~\cite{Kirkwood:1939br, frohlich1958theory}.  With the development of molecular dynamics methods, such hamiltonians have also been used for finite field simulations. The electric field in the coupling term is traditionally the applied electric field. Polar liquids are extended systems as are solids. The key innovation brought about by the modern theory of polarization in solids\cite{King-Smith:1993prb,Resta:1994rmp,Resta:2007ch} is replacing the applied electric field by the macroscopic electric field which includes the internal field generated by the polarization\cite{Stengel:2009cd}. This, from the perspective of physical chemistry, rather bold move was made by Stengel, Spaldin and Vanderbilt(SSV) on the basis of a clear understanding of what is included in Ewald summation of electrostatic interactions in periodic systems and what is not. This also gave the theory a firm foundation in the thermodynamics of dielectrics\cite{Landau:1960v8} which enabled SSV to transform their constant $\mathbf{E}$ to a constant $\mathbf{D}$ Hamiltonian.  

This study is a report on the application of the finite temperature classical force field variant of the SSV scheme in a calculation of the dielectric constant of SPC/E liquid water.  We started by rederiving the established fluctuation expression of the dielectric constant for supercells which is generally credited to deLeeuw-Perram-Smith\cite{DeLeeuw:1980rs80a,DeLeeuw:1980rs80b} and Neumann\cite{Neumann:1983kg,Neumann:1984mp}. Coupling polarization directly to the macroscopic electric field made this derivation a simple exercise in perturbation theory. A system under $\mathbf{E}=0$ constraints (equivalent to standard Ewald summation) can be interpreted as a short circuited capacitor (Fig.~\ref{fig0}a). Using the new SSV constant $\mathbf{D}$ hamiltonian we also obtained the corresponding fluctuation formula for the dielectric constant under $\mathbf{D}=0$ conditions which can be compared to open circuit conditions (Fig.~\ref{fig0}b). The theory was tested in  $\mathbf{E}=0$ and  $\mathbf{D}=0$ molecular dynamics simulation. Complementary finite $\mathbf{E}$ and $\mathbf{D}$ simulations were carried out to compare to the dielectric constant calculated directly from the field derivatives. The estimate of the dielectric constant obtained from  $\mathbf{E}=0$ and $\mathbf{D}=0$ polarization fluctuations were found to be in good agreement with each other and with the estimates from finite field derivatives, validating the SSV method for classical force field based MD. 

The motivation for this study was the possibility that application of constant $\mathbf{D}$ methods could reduce the costs of the computation of the static dielectric constant $\epsilon$. This expectation was based on the theory of dielectric relaxation predicting that decay of longitudinal polarization is significantly faster compared to transverse polarization. Arguing that polarization fluctuations under constant $\mathbf{D}$ are longitudinal, one can hope that  fixing the dielectric displacement instead of the electric field will speed up the convergence of averages and second moments of polarization. 

This prediction was verified by the simulation. The significant gain in time scale turned out to be however of limited help for the computation of $\epsilon$ from fluctuations at $\mathbf{D}=0$ because of the more stringent demands on accuracy. The reason is that $\epsilon$  under constant electric displacement must be computed from the inverse of a small number depending on the fluctuations(Eq.~\ref{epsd}). The calculation of the dielectric constant from electric displacement derivatives suffers in principle from a similar problem(Eq.~\ref{epsd2}). Fortunately the statistics in this case is more favourable. A further advantage is that non-linear effects in the response to a change in displacement field are very modest compared to a change in the electric field making the extrapolation to zero field easier. The constant $\mathbf{D}$ method may be therefore in the end the best option for the computation of dielectric response in DFT based MD, which is, for water, still effectively out of reach when using zero or constant $\mathbf{E}$ methods.

We conclude with an outlook. The modern theory of polarization was developed to resolve the fundamental question of the treatment of polarization in solids. The founding fathers of the theory of polar liquids had no such problems, using a system consisting of point dipoles as their basic model.\cite{Kirkwood:1939br,Onsager:1936gq} However, this simplicity is lost for realistic point charge models of polar molecules  which led to rather confusing arguments about the contribution of  higher multipole moments to the polarization\cite{Templeton:2013jcp}. This issue is avoided in the modern of polarization by a strict focus on macroscopic polarization, at the expense of turning polarization into a multivalued quantity.\cite{King-Smith:1993prb,Resta:1994rmp,Resta:2007ch}  

The multivalued polarization could be ignored in the present application to liquid water. We simply held on to polarization as the sum of molecular dipole moments viewing it as a special gauge appropriate for molecule systems. This will no longer work for applications to interfaces between solids and electrolytic solutions. Liquid-solid interfaces are described in the framework of macroscopic Maxwell theory by dividing the system up in piece-wise uniform dielectric continua. This conventional approach is regarded by some as incompatible with microscopic theory\cite{Matyushov:2014jcp}. To resolve these problems it would be useful to extend the modern theory of polarization by reintroducing some form of local polarization. The question of local polarization has already been addressed in the context of modelling of solid-solid heterojunctions\cite{Stengel:2009prb,Stengel:2009prr,Stengel:2011prl,Giustino:2005prb}. It remains to be seen whether these concepts can be applied to the electrical double layers formed at electrolyte-solid interfaces. If anything, the challenges in this area of research, and physical electrochemistry in general, should be an inspiration for both physical chemists and solid state physicists and we are hopeful that progress can be made in the near future.

\begin{acknowledgments}
Research fellowship (No.~ZH 477/1-1) provided by German Research Foundation (DFG) for CZ is gratefully acknowledged.  CZ thanks for helpful discussions with P.~Wirnsberger on the modified Green function approach to Ewald summation. CZ and MS also thank R.~M.~Lynden-Bell for encouraging discussions. We are in particular grateful to A.~Maggs for his explanation of an enlightening alternative view of polarization in periodic systems and to O.~Steinhauser for his clarification of reaction field methods.
\end{acknowledgments}

\appendix

\section{Legendre transforms} \label{sec:leg}
From the way they appear in the electric work relation Eq.\ref{ework}, the macroscopic field $\mathbf{E}$ and dielectric displacement $\mathbf{D}$ must be considered as thermodynamic conjugate variables\cite{Landau:1960v8}. This suggests that the electric enthalpy $F(\mathbf{E})$ of Eq.~\ref{fmd} and internal energy  $U(\mathbf{D})$ of Eq.~\ref{umd} are each others Legendre transform. However, while according to Eq.~\ref{dudd} the $\mathbf{D}$ derivative of $U$ yields $\mathbf{E}$, Eq.~\ref{dfde} is not consistent with a Legendre transform. The $\mathbf{E}$ derivative of $F$ is not reproducing  $-\mathbf{D}$.  Following Landau and Lifshitz SSV adjust the definition of $F(\mathbf{E})$ to\cite{Landau:1960v8}
\begin{equation}
\label{ftilde}
\tilde{F}(\mathbf{E}) = {F}(\mathbf{E}) - \frac{\mathit{\Omega}}{8\pi} \mathbf{E}^2
\end{equation}
Then with Eq.~\ref{ddef} 
\begin{equation}
\label{dtilde}
 \frac{d \tilde{F}}{d \mathbf{E}} = - \frac{\mathit{\Omega}}{4 \pi} \mathbf{D}
\end{equation}
as required. The term added to $F(\mathbf{E})$ is a constant in the ensemble generated by the electric enthalpy Hamiltonian of Eq.~\ref{fhmd} and will therefore not affect averages over the ensemble.      

Writing $\tilde{F}$ and $U$ in thermodynamic form as the sum  of an internal energy and entropic ($-TS$) term we find the expected Legendre transform relation
\begin{equation}
\label{legend}
\tilde{F}(\mathbf{E}) = \mathrm{U}(\mathbf{D}) - \frac{\mathit{\Omega}}{4\pi} \mathbf{E}\cdot \mathbf{D}
\end{equation} 
$U$  can therefore be equated with the thermodynamic potential with respect to $\mathbf{D}$ and $\tilde{F}$ as the thermodynamic potential with respect to $\mathbf{E}$ referred to by Landau and Lifshitz as $U$ respectively $\tilde{U}$\cite{Landau:1960v8}. The negative sign of the  $\mathit{\Omega} \mathbf{E}^2/8 \pi$ term in Eq.~\ref{ftilde} is consistent with this interpretation. The sign of field energy in electric enthalpy is opposite (negative) to the sign in internal energy\cite{Landau:1960v8}. Subtracting $\mathit{\Omega} \mathbf{E}^2/8 \pi$ in Eq.~\ref{ftilde} therefore amounts to adding in the energy of the constant macroscopic field as explained in the supporting information of Ref.~\citenum{Stengel:2009cd}

SSV also consider the Legendre transform of $F(\mathbf{E})$ with respect to $\mathbf{P}$ using Eq.~\ref{dfde} leading to the thermodynamic potential
\begin{equation}
\label{epfix}
  E (\mathbf{P}) = F(\mathbf{E}) + \mathit{\Omega} \mathbf{E} \cdot \mathbf{P}  
\end{equation}
with the polarization derivative
\begin{equation}
\label{dedp}
 \frac{dE}{d\mathbf{P}} = \mathit{\Omega} \mathbf{E}  
\end{equation}
The statistical mechanics generated by this potential corresponds to an ensemble at fixed polarization $\mathbf{P}$\cite{Stengel:2009cd} which was used in the pioneering studies of the electrical equation of state of ferroelectric materials\cite{Sai:2004prb,Dieguez:2006prl}. 

The function $E(\mathbf{P})$ also has already a long history in physical chemistry\cite{Alberty:2001pac} (going back to gas-phase chemical thermodynamics). This is presumably the reason why it was chosen by Aragones et al. as the starting point in their study of the electric field dependence of the phase diagram of ice\cite{Vega:2011prl}. They then change this into a constant $\mathbf{E}$ approach by applying the reverse Legendre transform giving them $F(\mathbf{E})$ and the corresponding electric enthalpy Hamiltonian of Eq.~\ref{fhmd}.  Finally to convert to a constant applied field method the macroscopic field $\mathbf{E}$ in Eq.~\ref{fhmd} is replaced by $\mathbf{E}_0$. Aragones et al. base their approach on the thermodynamic theory of Landau and Lifshitz. Their finite field method is consistent with the theory in section \ref{sec:EDham} which provides a further microscopic basis and a clarification of the distinction between the applied and macroscopic field (see section \ref{sec:physchem}).     

\section{Surface terms} \label{sec:surf}
The issue of surface terms arose when it is was realized that the electrostatic interactions in a finite but large ionic crystal can be modelled by an intrinsic energy equal to the ``tinfoil'' Ewald energy of a supercell in the infinite crystal and a shape dependent extrinsic term\cite{Redlack:1975jpcs}. The derivation familiar to physicists and chemist alike is due to deLeeuw, Perram and Smith (LPS)\cite{DeLeeuw:1980rs80a,DeLeeuw:1980rs80b}. The problem was revisited again and again generating an extensive literature from which we only quote a small subset\cite{Smith:1981rsa,Smith:2008jcp,Caillol:1994jcp,Foulkes:1996prb,Kantorovich:1999jpcm,Ballenegger:2014jpc}. SSV add two field dependent coupling terms to the Ewald Hamiltonian.
From Eq.~\ref{fhmd} we have for constant $\mathbf{E}$
\begin{equation}
\label{ve}
  V_{\mathbf{E}} = -\Omega \mathbf{E} \cdot \mathbf{P}
\end{equation}
and for constant $\mathbf{D}$ from Eq.~\ref{uvdb}
\begin{equation}
\label{vd}
  V_{\mathbf{D}} = \frac{\Omega}{8 \pi} 
  \left(\mathbf{D} - 4 \pi  \mathbf{P}\right)^2
\end{equation}  
The question investigated in this appendix is whether $V_{\mathbf{E}}$ and $V_{\mathbf{D}}$ can be regarded as a surface terms. To this end we consider a piece of dielectric material with an homogeneous polarization density $\mathbf{P}$ representing the $\mathbf{k}=0$ component of the instantaneous polarization density in a liquid. Terminating the dielectric at an interface creates a polarization surface charge density $\sigma_p$  
\begin{equation}
\label{sigmap}
\sigma_p = -\mathbf{n} \cdot  \mathbf{P} 
\end{equation}
where $\mathbf{n}$ is the normal to the bounding surface pointing inward. $\sigma_p$  generates an electric field called the polarization field
\begin{equation}
\label{ep}
\mathbf{E}_p = - \nabla_{\mathbf{r}} \int dA \frac{ \sigma_p(\mathbf{r}^\prime)}{| \mathbf{r} - \mathbf{r}^\prime |}
\end{equation}
In the context of the physics of ferroelectricity $\mathbf{E}_p$ is usually referred to as the depolarizing field. $\mathbf{E}_p$ is closely related to the longitudinal component  of the polarization, which will be defined below. 

The separation into a  longitudinal component $\mathbf{P}_L$ and transverse component $\mathbf{P}_T$ is a rigorous mathematical result of the Helmholtz theorem of vector calculus\cite{Lehrer:2010sp}
\begin{equation}
\label{plt}
\mathbf{P}   = \mathbf{P}_L + \mathbf{P}_T 
 \end{equation}
 with $ \mathbf{P}_L$ and  $ \mathbf{P}_T$ given
 \begin{eqnarray}
\label{pl}
 \mathbf{P}_L (\mathbf{r}) & = &  - \nabla_{\mathbf{r}} \phi  (\mathbf{r})  \\[6pt]
   \phi  (\mathbf{r}) & = &   \int_V d\mathbf{r}^\prime
   \frac{ \nabla_{\mathbf{r}^\prime} \cdot \mathbf{P}(\mathbf{r}^\prime)}
  {4 \pi | \mathbf{r} - \mathbf{r}^\prime |} 
  + \int_A dA \frac{\mathbf{n}(\mathbf{r}^\prime) 
\cdot \mathbf{P}(\mathbf{r}^\prime)} {4 \pi | \mathbf{r} - \mathbf{r}^\prime |}
  \nonumber \\[6pt]
 \mathbf{P}_T (\mathbf{r}) & = & \nabla_{\mathbf{r}} \wedge \mathbf{Q}  (\mathbf{r}) 
 \label{pt} \\[6pt]
  \mathbf{Q}  (\mathbf{r}) & = & \int_V d\mathbf{r}^\prime
   \frac{ \nabla_{\mathbf{r}^\prime} \wedge \mathbf{P}(\mathbf{r}^\prime)}
  {4 \pi | \mathbf{r} - \mathbf{r}^\prime |} 
 -  \int_A dA \frac{\mathbf{n}(\mathbf{r}^\prime) \wedge 
\mathbf{P}(\mathbf{r}^\prime)}{4 \pi | \mathbf{r} - \mathbf{r}^\prime |} 
\nonumber
\end{eqnarray}
so that  $\nabla \wedge \mathbf{P}_L  = 0$ and $\nabla \cdot \mathbf{P}_T = 0$. The reader is referred to Matyushov for a discussion of the role of transverse polarization in solvation and hydration\cite{Matyushov:2014jcp,Matyushov:2004jcp}.  

For homogeneous polarization, there are no volume contributions, only surface terms.  The surface terms are in principle position dependent, even if their sum $\mathbf{P}$ is homogeneous. However, because $\mathbf{P}$ is constant over the whole of the body, $\nabla \cdot \mathbf{P}_L = \nabla \cdot \left( \mathbf{P} - \mathbf{P}_T \right) = 0 $. Similarly $\nabla \wedge \mathbf{P}_T = 0$. $\mathbf{P}_L$ and $\mathbf{P}_T$ are cavity fields, satisfying vacuum electrostatic equations. A possible $\mathbf{r}$ dependence of $\mathbf{P}_L$ and $\mathbf{P}_T$ will be henceforward suppressed.

Comparing Eqs.~\ref{sigmap}, \ref{ep} and \ref{pl}  we obtain a general equation for the depolarizing field of a homogeneously polarized dielectric body of arbitrary shape identifying it with the longitudinal component of polarization.  
\begin{equation}
\label{eppl}
  \mathbf{E}_p = - 4 \pi \mathbf{P}_L
\end{equation}
Adding the applied field $\mathbf{E}_0$ gives the macroscopic field 
\begin{equation}
\label{e0ep}
 \mathbf{E} = \mathbf{E}_0 + \mathbf{E}_p 
\end{equation}
and therefore with Eq.~\ref{eppl}
\begin{equation}
\label{e0pl}
\mathbf{E} = \mathbf{E}_0 - 4 \pi \mathbf{P}_L
\end{equation}
Next we reformulate Eqs.~\ref{e0ep} and \ref{e0pl} replacing the applied electric field $\mathbf{E}_0$ by the more fundamental displacement field $\mathbf{D}$ defined by the relation
\begin{equation}
\label{dep}
  \mathbf{D} = \mathbf{E} + 4 \pi \mathbf{P}
\end{equation}
Substituting Eq.~\ref{plt} and \ref{e0pl} we find
\begin{equation}
\label{de0pt}
  \mathbf{D} = \mathbf{E}_0 + 4 \pi \mathbf{P}_T
\end{equation}
confirming that the longitudinal component $\mathbf{D}_L$  of the electrostatic induction $\mathbf{D}$ can be treated as an applied electric field $\mathbf{E}_0$.  
However, Eq.~\ref{de0pt} also states that electrostatic induction cannot be simply equated to the applied field. Depending on the shape of the bounding surface $\mathbf{D}$ may contain a transverse residue.

The $\mathbf{E}=0$ system is the original ``tinfoil boundary geometry'' of LPS. There is no surface term, and also $V_E$ of Eq.~\ref{ve} is zero for zero field. This is straight forward. The difficulty is the  $\mathbf{D} =0 $ system. For zero displacement field the function $V_D$ of Eq.~\ref{vd} becomes equal to
\begin{equation}
\label{vd0}
  V_D(0) = 2 \pi\Omega \mathbf{P}^2
 \end{equation}
Can this term can be interpreted as a LPS-type surface term? The answer to this question is negative.  We base our argument on work by Kantorovitch\cite{Kantorovich:1999jpcm}, who showed that the surface term of an ellipsoid shaped cluster of supercell replicas can be written in the form
 \begin{equation}
\label{vs}
 V_s \left(\mathbf{E}\right)= - \frac{\Omega}{2} \mathbf{E}_p \cdot \mathbf{P}
\end{equation}
where $\mathbf{E}_p$ is the depolarizing field given by Eq.~\ref{ep}. Recently 
Ballenegger showed that this relation is valid for a smooth surface of arbitrary shape\cite{Ballenegger:2014jpc}. The dependence on surface geometry is implicit in the  polarization field $\mathbf{E}_p$. Clearly $V_D(0)$ of Eq.~\ref{vd0} and $V_s$ of Eq.~\ref{vs} are equal if
\begin{equation}
\label{epp}
   \mathbf{E}_p = - 4 \pi \mathbf{P}
\end{equation} 
Comparing to Eq.~\ref{eppl} we see that for a geometry to satisfy Eq.~\ref{epp} the polarization must be entirely longitudinal $\mathbf{P}= \mathbf{P}_L$ or $\mathbf{P}_T = 0$. This can be realized in special directions. For a isotropically fluctuating polarization it must hold for all directions.  The dielectric response considered by LPS is isotropic. However, for a sphere $\mathbf{P}_L = \mathbf{P}/3$  or equivalently $\mathbf{P}_T = 2 \mathbf{P}_L$  (see for example Ref.~\citenum{Lehrer:2010sp}). The depolarizing field of a sphere fails Eq.~\ref{epp}. 

Do closed surfaces for which $\mathbf{P}_T$ is strictly zero exist? As far as we are aware they don't. There will always be some direction in which the depolarizing field is less than what the full polarization would give (Eq.~\ref{epp}). This also implies that there is no dielectric body for which the displacement field is equal to a finite applied electric field of arbitrary orientation (Eq.~\ref{de0pt}). We are unable to provide a mathematical proof for this statement which must therefore remain a hypothesis.  This hypothesis is however strongly supported by the work of Maggs who pointed out that Eq.~\ref{epp} is in fact satisfied for a periodic system \emph{without} surfaces\cite{Maggs:2004jcp,Maggs:2006prl}. While such a geometry cannot be realized experimentally, it can be considered as a theoretical and computational construction. The argument is briefly summarized below. For further justification and discussion we refer to the original papers. 

Maggs starts from the Helmholtz theorem Eqs.~\ref{plt}-\ref{pt}, but now applied to a general inhomogeneous electric field $\mathbf{E}(\mathbf{r})$. 
\begin{equation}
\label{maggs1}
  \mathbf{E}(\mathbf{r}) = - \nabla \phi(\mathbf{r}) + \nabla \wedge \mathbf{Q}(\mathbf{r}) + \bar{\mathbf{E}} 
\end{equation}
 The first term is the longitudinal field $\mathbf{E}_L = -\nabla \phi$ derived from a scalar potential $\phi$.  The formalism also allows for a transverse component $\mathbf{E}_T =  \nabla \wedge \mathbf{Q}$. The system is periodic, there are no  boundaries generating surface terms. This is the crucial difference with LPS approach. Instead we have in Eq.~\ref{maggs1} the $\mathbf{r}$ independent  term $\bar{\mathbf{E}}$, which is an as yet unspecified uniform field. 

Using an infinite system from the start disregarding surfaces is also the basis of the derivation of the frequency and wavevector dependent dielectric and magnetic linear response functions by Fulton\cite{Fulton:1975ii} and Madden and Kivelson\cite{Madden:1984cp}. The crucial step proposed by Maggs is to link the uniform ($\mathbf{k} =0$) electric field  $\bar{\mathbf{E}}$ to the polarization using the  time dependent Maxwell equation
\begin{equation}
\label{maggs2}
  \frac{ \partial \mathbf{E}(t)}{\partial t} = - 4 \pi \mathbf{J} + c \nabla \wedge \mathbf{B}
\end{equation}
where $\mathbf{J}$ is the current and $\mathbf{B}$ the magnetic field. $c$ is the velocity of light. Eq.~\ref{maggs2} is a microscopic Maxwell equation in Gaussian units. Integrating over time and space gives the uniform polarization
\begin{equation}
\label{maggs3}
  \bar{\mathbf{E}}(t) = - \frac{4 \pi}{\Omega} \int_{t_0}^t dt^\prime \int_{cell} d\mathbf{r} \, \mathbf{J}(t^\prime) 
   \equiv - 4 \pi \mathbf{P}(t) 
\end{equation}
where we have used that the spatial integral of a curl  over the unit cell of a periodic system vanishes\cite{Maggs:2004jcp}. We can imagine that the system was subjected to some action starting  from an unpolarized reference state at time $t_0$. The adiabatic current this action creates establishes a final polarized state. This is also how polarization is defined in the modern theory of polarization.  
Eq.~\ref{maggs3} relates the homogeneous polarization $\mathbf{P}$  to an internal electric field $\bar{\mathbf{E}}$. This the only electric field there is, hence with Eq.~\ref{dep} we conclude that $\mathbf{D}=0$ which then allows us to equate $\bar{\mathbf{E}}$ of Eq.~\ref{maggs3} to the depolarizing field $\mathbf{E}_p$ satisfying Eq.~\ref{epp}. The electric energy of a volume $\mathit{\Omega}$ cut out of the system is $\mathit{\Omega}\mathbf{E}^2/8 \pi$. Replacing the filed by the polarization and adding the Ewald sum representing all $\mathbf{k} \neq 0$ contributions to the energy leads to the $\mathbf{D}=0$ SSV Hamiltonian of Eq.~\ref{u0vdb}.  
    
The time dependent scheme proposed by Maggs and the modern theory of polarization have much in common. Both schemes use the current to define polarization in a periodic system without specifying a surface.  This applies to $\mathbf{D}=0$ boundary conditions as well, which are normally associated with a finite body in vacuum. This may seem counterintuitive, but is consistent with the geometry invariance of the constitutive relations\cite{Landau:1960v8}.  

\section{Reaction fields} \label{sec:reac}
In parallel to the Ewald summation based methods of the deLeeuw, Perram and Smith, an alternative approach based on reaction field methods was developed by Neumann and Steinhauser(NS)\cite{Neumann:1983gy,Neumann:1983gz,Neumann:1983kg}. The central result is a general fluctuation expression for the static dielectric constant $\epsilon$ of a spherical body of polar fluid embedded in a dielectric continuum.  
\begin{equation}
\label{flucns}
\frac{\epsilon-1}{\epsilon+2} = \frac{4 \pi}{3}
 \left(\frac{\beta \Omega \langle \mathbf{P}^2 \rangle}{3} \right)
 \left[1-\frac{3}{4 \pi}\frac{\epsilon-1}{\epsilon+2} T_{mod}(\epsilon_R)\right]
\end{equation}
Following the notation of section \ref{sec:fluc}, $\mathbf{P}$ is the polarization computed from the total dipole moment of a system of volume $\Omega$. The factor $T_{mod}(\epsilon_R)$ is the volume average of the dipolar field tensor adapted (``modified'') for interaction with the embedding dielectric continuum of dielectric constant $\epsilon_R$. $T_{mod}(\epsilon_R)$ is closely related to the Onsager reaction field of a dipole\cite{frohlich1958theory}. 
\begin{equation}
\label{reacns}
 T_{mod}(\epsilon_R)=
 \frac{4 \pi}{3}\frac{2 \left(\epsilon_R -1 \right)}{2 \epsilon_R + 1 }
\end{equation}
For the derivation of Eqs.~\ref{flucns} and \ref{reacns} we refer to the original papers by NS\cite{Neumann:1983gy,Neumann:1983kg}. Substituting $\epsilon_R = \epsilon$ in Eq.~\ref{flucns} recovers the Kirkwood-Fr\"ohlich expression for the dielectric constant\cite{Neumann:1983kg}. However, $\epsilon_R$ can also be different from $\epsilon$. While this changes the magnitude of the polarization fluctuations, Eq.~\ref{flucns} combined with the reaction field factor Eq.~\ref{reacns} still gives the correct relation to the dielectric constant. In particular, the limit  $\epsilon_R \rightarrow \infty$ corresponds to a polar fluid in a spherical cavity in a metal. The reaction field factor Eq.~\ref{reacns} becomes 
\begin{equation}
\label{reacinf}
 T_{mod}(\infty)=  \frac{4 \pi}{3}
\end{equation}
which inserted in Eq.~\ref{flucns} yields
\begin{equation}
\label{epsinf}
\epsilon-1 = \frac{4 \pi \beta \Omega}{3} \langle \mathbf{P}^2 
  \rangle_{\epsilon_R=\infty}
\end{equation}
Eq.~\ref{epsinf}  is identical to the relation of Eq.~\ref{epse} between $ \langle \mathbf{P}^2  \rangle_{\mathbf{E}=0}$ and the dielectric constant (for simplicity here we have assumed that $\langle \mathbf{P} \rangle = 0 $ as it should be in a converged MD run).  This led NS to identify Ewald summation with their $\epsilon_R = \infty$ reaction field limit\cite{Neumann:1983gy,Neumann:1983gz,Neumann:1983kg}

The opposite case of a sphere in vacuum is obtained by setting $\epsilon_R = 1$, and hence  
\begin{equation}
 T_{mod}(1)= 0
\end{equation}
giving the fluctuation relation
\begin{equation}
\label{flucvac}
\frac{\epsilon-1}{\epsilon+2} = \frac{4 \pi \beta \Omega}{9} \langle \mathbf{P}^2 \rangle_{\epsilon_R=1}
\end{equation}
 As can already be expected from the argument of appendix \ref{sec:surf}, the vacuum equation \ref{flucvac} is not equal to the fluctuation relation Eq.~\ref{epsd} under $\mathbf{D}=0$ conditions.  

However, as noticed by Caillol and coworkers\cite{Caillol:1994jcp,Caillol:1989jcp} the NS reaction field approach is in fact capable of reproducing the $\mathbf{D}=0$ Hamiltonian Eq.~\ref{u0vdb} (see also Ref.\citenum{Maggs:2004jcp}). This is achieved by setting the dielectric constant of the embedding medium to the ``unphysical'' value of $\epsilon_R=0$. Inserting in Eq.~\ref{reacns} gives 
\begin{equation}
 T_{mod}(0)= - \frac{8 \pi}{3}
\end{equation}
which should be compared to Eq.~\ref{reacinf} valid for $\mathbf{E}=0$ (note the change of sign). Substituting in Eq.~\ref{flucns} we find
\begin{equation}
\label{dns}
\frac{\epsilon-1}\epsilon = \frac{4 \pi \beta \Omega}{3} 
 \langle \mathbf{P}^2 \rangle_{\epsilon_R = 0}
\end{equation}
which is indeed the counterpart to Eq.~\ref{epsd}. Similarly,  dividing 
Eq.~\ref{epsinf} by Eq.~\ref{dns} gives the same ratio as Eq.~\ref{ratio}
\begin{equation}
\frac{\langle \mathbf{P}^2\rangle_{\epsilon_R=\infty}}
{ \langle \mathbf{P}^2 \rangle_{\epsilon_R = 0}} = \epsilon
\end{equation}
confirming that $\mathbf{D}=0$ boundary conditions are contained in the NS reaction field formalism, all be it for an environment with the unphysical dielectric constant of zero. 

The interpretation of this rather surprising connection is better left to the experts in reaction field methods. As mentioned it was noticed later and is not discussed in the original papers on the NS method. We furthermore point out, that the thermodynamic perspective underlying the SSV method greatly simplifies the derivation of the zero field fluctuation relations avoiding the complication of the dipolar tensor. Moreover, it allows for natural extension to finite displacement fields $\mathbf{D}$ as we have shown in the present paper.


%

\end{document}